\documentclass[journal,onecolumn]{IEEEtran}

\usepackage{cite}

\usepackage{amsmath}

\usepackage{tabularx}
\newcolumntype{Y}{>{\centering\arraybackslash}X}
\def\arraystretch{1.2}%

\usepackage{url}
\usepackage{hyperref}
\hypersetup{
    colorlinks=true,
    linkcolor=blue,
    filecolor=magenta,      
    urlcolor=cyan,
}

\usepackage{graphicx}
\graphicspath{ {./figures/} }

\usepackage{csquotes}
\usepackage{caption}
\usepackage{longtable}
\usepackage{float}
\usepackage{tikz}
\def\checkmark{\tikz\fill[scale=0.4](0,.35) -- (.25,0) -- (1,.7) -- (.25,.15) -- cycle;}

\begin{document}

\title{Examination of tools for managing different dimensions of Technical Debt}

\author{Dwarak Govind Parthiban$^{1}$, University of Ottawa
\thanks{$^{1}$dpart016@uottawa.ca}
}

\maketitle

\begin{abstract}
With lots of freemium and premium, open and closed source  software tools that are available in the market for dealing with different activities of Technical Debt management across different dimensions, identifying the right set of tools for a specific activity and dimension can be time consuming. The new age cloud-first tools can be easier to get onboard, whereas the traditional tools involve a considerable amount of time before letting the users know what it has to offer. Also, since many tools only deal with few dimensions of Technical Debt like Code and Test debts, identifying and choosing the right tool for other dimensions like Design, Architecture, Documentation, and Environment debts can be tiring. We have tried to reduce that tiring process by presenting our findings that could help others who are getting into the field of \enquote{Technical Debt in Software Development} and subsequently further into \enquote{Technical Debt Management}. 
\end{abstract}

\IEEEpeerreviewmaketitle

\section{Introduction}

\IEEEPARstart{T}{echnical Debt} (TD) is a term that was coined by Ward Cunningham \cite{wiki} in the year of 1992. It is a concept in software development that reflects the implied cost of additional rework caused by choosing an easy solution now instead of using a better approach that would take longer. There are several dimensions of technical debt like code debt, test debt, documentation debt, environment Debt, design Debt, and architecture Debt. As with financial debt,  technical debt must be paid back, and is comprised of two parts: principal and interest.  In the software development metaphor, the interest is paid in the form of additional work required to maintain the software system given its sub-optimal code. Time spent improving the code, which isn’t directly adding customer value and which wouldn’t be necessary if the code were optimally designed currently, represents paying down the principal on the debt. \\
\vspace{2pt} \\
\IEEEPARstart{M}{anaging} the technical debt mostly consists in \enquote{repaying the principal} to achieve business value. Technical debt management is the set of activities that: prevent potential technical debt from being accrued, deal with the accumulated technical debt to make it visible and controllable, keep a balance between the cost and value of the software. Technical debt management includes the following eight activities: representation, prevention, communication, prioritization, monitoring, measurement, identification, and repayment of technical debt.  \\
\vspace{2pt} \\
\IEEEPARstart{IN}{this paper}, we first introduce a wide-array of software tools that are currently available in the market according to our knowledge. We then choose some among them which we believe to be complementary to each other, examine them further using some of the open source projects available in GitHub, and report our findings which is a subset of all the functionalities that those tools have. We present our findings in a such a way that the readers can associate them to the appropriate dimension of the technical debt and the corresponding activity within technical debt management. Finally, we propose a cost model for TD prinicipal calculation and suggest a new tool that has recently hit the market by providing details about it. 

\section{Tools and Projects}

\begin{table}[h!]
  \centering
  \begin{tabularx}{\columnwidth}{@{}|l|Y|Y|Y|Y|@{}}
   \hline
   \textit{Name} & \textit{Cloud} & \textit{On-premise} & \textit{Distribution} & \textit{Dimensions} \\
   \hline
   Sonarqube & & \checkmark & \href{https://github.com/SonarSource/sonarqube}{Open} & Code, Test\\
   \hline
   Sonarcloud & \checkmark & & Closed & Code, Test\\
   \hline
   Teamscale & & \checkmark & Closed & Code, Test, Documentation, Architecture\\
   \hline
   Codescene & \checkmark & & Closed & Code, Architecture\\
   \hline
   Codacy & \checkmark & & Closed & Code\\
   \hline
   DesigniteJava & & \checkmark & \href{https://github.com/tushartushar/DesigniteJava}{Open} & Code, Design\\
   \hline
   Scrutinizer & \checkmark & & Closed & Code\\
   \hline
   Lattix & & \checkmark & Closed & Design, Architecture\\ 
   \hline
   Jacoco & \checkmark (Sonarcloud plugin) & \checkmark (Standalone, Maven Plugin, Eclipse Plugin) & \href{https://github.com/jacoco/jacoco}{Open} & Test\\ 
   \hline
   Checkstyle & & \checkmark (Eclipse Plugin) & \href{https://github.com/checkstyle/checkstyle}{Open} & Code\\
   \hline
\end{tabularx}
     \vspace{5pt}

    \caption{Some of the tools available in the market}
    \label{tab:1}
\end{table}

Among the tools that are mentioned in table \ref{tab:1}, we chose Sonarcloud, Checkstyle, Jacoco, DesigniteJava, Codescene and Lattix. The reason for our selection is we wanted the tools to be mostly complementary to each other (in their default settings with no or very few customizations) so that we can cover many dimensions within technical debt than just focussing on code and test debts. Also, we wanted them to be appropriate for different activities within TD management.  

We also chose four open source Java projects which could be built using Maven. The chosen projects were used for the examination of the abovementioned tools. Some details about those projects are mentioned in table \ref{tab:2}. The measures: lines of code and number of classes were retrieved from Sonarcloud. The github stars takes into account of the total number of stars till April 15, 2019 12.30 PM EDT. 

\begin{table}[h!]
  \centering
  \begin{tabularx}{\columnwidth}{@{}|l|l|Y|Y|Y|Y|@{}}
   \hline
   \textit{Name} & \textit{Description} & \textit{Lines of Code} & \textit{Number of Classes} & \textit{GitHub Stars}  \\
   \hline
    \href{https://github.com/TooTallNate/Java-WebSocket}{Java WebSocket} & Barebones websocket client and server implementation & 5K & 65 & 5197 \\
    \hline
    \href{https://github.com/jankotek/JDBM3}{JDBM3} & Embedded Key Value Java Database & 9.1K & 63 & 312 \\
    \hline
    \href{https://github.com/xetorthio/jedis}{Jedis} & A blazingly small and sane redis java client & 20.8K & 129 & 7933 \\
    \hline
    \href{https://github.com/mybatis/mybatis-3}{MyBatis} & SQL mapper framework for Java & 22.6K &  397 &10,360 \\
   \hline
\end{tabularx}
     \vspace{5pt}

    \caption{Java projects used for examination}
    \label{tab:2}
\end{table}

\section{Findings}

\subsection{Quality assessment}

\begin{enumerate}
    \item Among the chosen tools, Sonarcloud and Lattix allow assessing the following software quality attributes, 
    \begin{itemize}
        \item Sonarcloud - Reliability, Maintainability, Security [See figures \ref{fig:rel}, \ref{fig:mai}, and \ref{fig:sec} respectively]
        \item Lattix - Stability [See figures \ref{fig:stab_jdb}, \ref{fig:stab_jws}, \ref{fig:stab_jed},  and \ref{fig:stab_myb}]
    \end{itemize}
    \item Below are the screenshots of quality attributes from the abovementioned tools, \par
    \begin{minipage}{0.9\columnwidth}
		  \centering
	     \includegraphics[height=0.45\textheight,keepaspectratio]{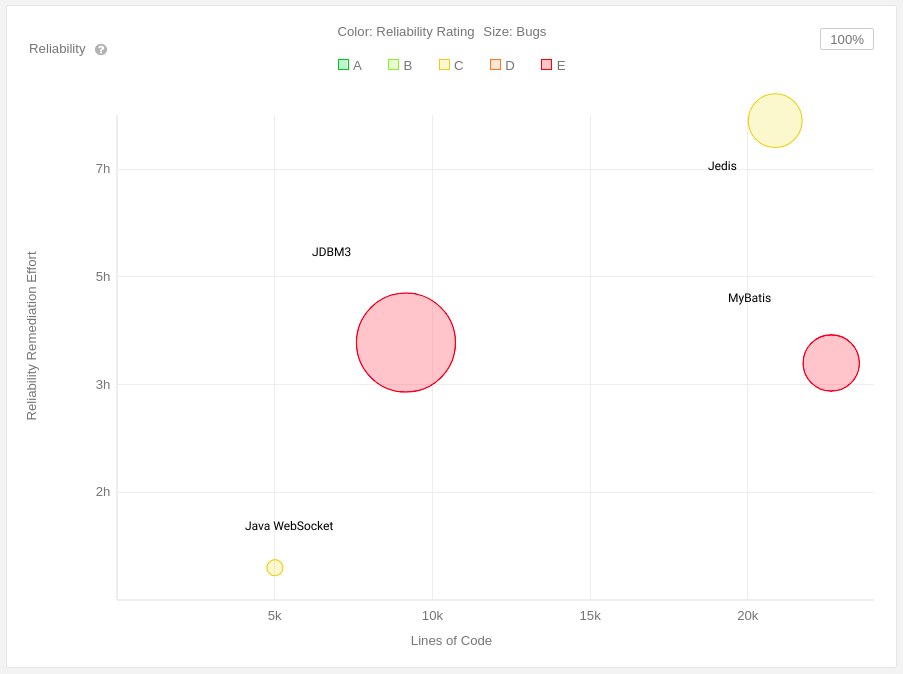}
	     \captionof{figure}{Reliability}
	     \label{fig:rel}
	\end{minipage}\\
	 \begin{itemize}
	 \item Reliability attribute helps in seeing bugs' operational risks to the projects. The closer a bubble's color is to red, the more severe the worst bugs in the project. Bubble size indicates bug volume in the project, and each bubble's vertical position reflects the estimated time to address the bugs in the project. Small green bubbles on the bottom edge are best.
	 \end{itemize} 
 	\begin{minipage}{0.9\columnwidth}
	  \centering
	   \includegraphics[height=0.45\textheight,keepaspectratio]{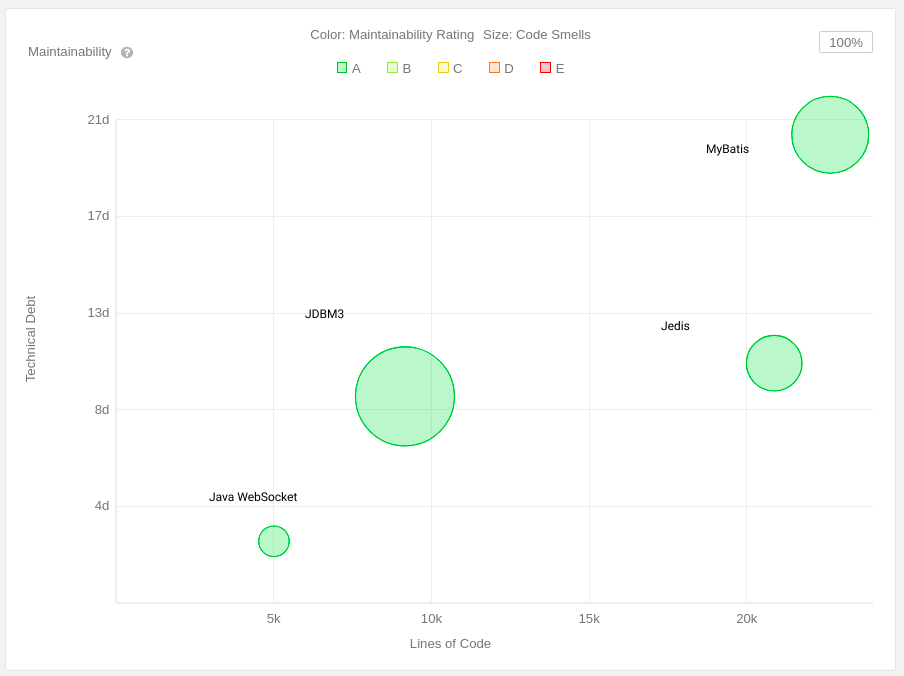}
	   \captionof{figure}{Maintainability}
	   \label{fig:mai}
	\end{minipage}\\
	\begin{itemize}
	\item Maintainability attribute helps in seeing code smells' long-term risks to the projects. The closer a bubble's color is to red, the higher the ratio of technical debt to project size. Bubble size indicates code smell volume in the project, and each bubble's vertical position reflects the estimated time to address the code smells in the project. Small green bubbles on the bottom edge are best.
	\end{itemize}
	\begin{itemize}
	\item Security attribute helps in seeing vulnerabilities' operational risks to your projects. The closer a bubble's color is to red, the more severe the worst vulnerabilities in the project. Bubble size indicates vulnerability volume in the project, and each bubble's vertical position reflects the estimated time to address the vulnerabilities in the project. Small green bubbles on the bottom edge are best.
	\end{itemize}
	\begin{itemize}
	\item Stability attribute of a system reports how sensitive the architecture is to the changes in atoms (say classes) within the subsystem. A higher stability value corresponds to less dependency on atoms within the selected system.
	\end{itemize}
	\begin{minipage}{0.9\columnwidth}
		\centering
		\includegraphics[keepaspectratio]{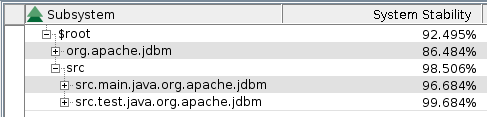}
		\captionof{figure}{System stability (overall and top-level components) for JDBM3}
		\label{fig:stab_jdb}
	\end{minipage}\\
  	\begin{minipage}{0.9\columnwidth}
		  \centering
	    \includegraphics[height=0.45\textheight,keepaspectratio]{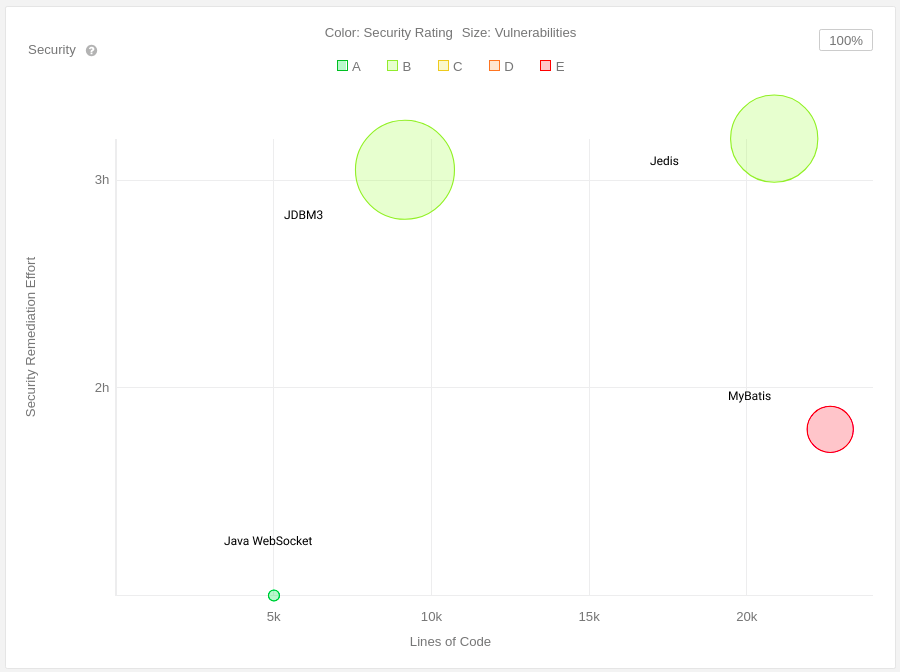}
	    \captionof{figure}{Security}
	    \label{fig:sec}
 	\end{minipage}\\
	\begin{minipage}{0.9\columnwidth}
		\centering
		\includegraphics[keepaspectratio]{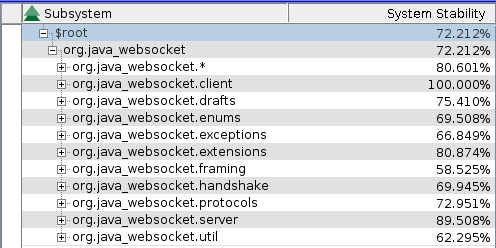}
		\captionof{figure}{System stability (overall and top-level components) for Java WebSocket}
		\label{fig:stab_jws}
	\end{minipage}\\

	\begin{minipage}{0.9\columnwidth}
		\centering
		\includegraphics[keepaspectratio]{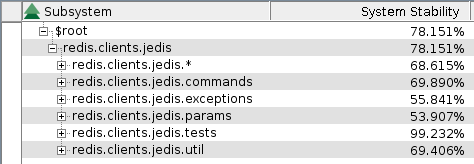}
		\captionof{figure}{System stability (overall and top-level components) for Jedis}
		\label{fig:stab_jed}
	\end{minipage}\\
	\begin{minipage}{0.9\columnwidth}
		\centering
		\includegraphics[keepaspectratio]{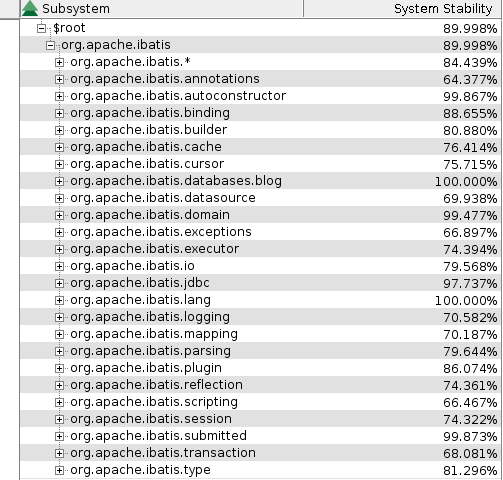}
		\captionof{figure}{System stability (overall and top-level components) for MyBatis}
		\label{fig:stab_myb}
	\end{minipage}\\

	\pagebreak
    \item Based on multiple quality assessments, \par
    \begin{table}[H]
      \centering
      \begin{tabularx}{0.9\columnwidth}{|Y|Y|Y|Y|Y|Y|Y|}
       \hline
       \textit{Project} & \textit{Reliability rank} &\textit{Maintainability rank} & \textit{Security rank} & \textit{Stability rank} & \textit{Sum} & \textit{Overall quality rank} \\
       \hline
       Java WebSocket & 1 & 1 & 1& 4& 7& 1\\
       \hline
       JDBM3 & 3 & 2 & 2& 1& 8& 2\\
       \hline
       Jedis & 2 & 3 & 3& 3& 11& 3\\
       \hline
       MyBatis & 3 & 4 & 4& 2& 13& 4\\
       \hline
    \end{tabularx}
    \caption{Overall quality rank}
    \label{tab:q_rank}
    \end{table}
    \begin{itemize}
    \item In the above table, the ranks for quality attributes are based on the rating given by the tools. For example, if project X and Y get security ratings of A and B, security rank of X is higher (1 is higher than 2) than that of Y. If two projects have same rating, then higher rank is given to the one who has less value in the Y axis, which is typically the time taken for remediation efforts.
    \item The overall quality rank [Column 6]  is calculated by ranking the sums [Column 5] of individual ranks [Columns 1, 2, 3, 4].  
    \end{itemize}
\end{enumerate}

\subsection{Technical debt identification}

\begin{table}[h!]
  \centering
  \begin{tabularx}{\textwidth}{|Y|Y|}
   \hline
   \textit{Dimension} & \textit{Some Items}  \\
   \hline
    Code debt & Coding guideline violations, Code smells, Inconsistent style\\
    \hline
    Design debt & Design rule violations, Design smells, Violation of design constraints\\
    \hline
    Test debt & Lack of tests, Inadequate test coverage, Improper test design \\
    \hline
    Architecture debt & Architecture rule violations, Modularity violations, Architecture smells \\
   
   \hline
\end{tabularx}
     \vspace{5pt}

    \caption{Categorization of some of the technical debt items associated with their technical debt dimension according to \cite{classification}}
    \label{tab:3}
\end{table}

With the use of these many tools, it is obvious that combinedly the tools can identify a lot of TD types with a lot more instances for each of those types. To describe a high-level diversity, we have just presented a subset of them in table \ref{tab:identified}. It could be seen that with current tools, the diversity of TD items in design and architecture dimensions seem to be lower than those present in code and test dimensions. 

\begin{table}[h!]
  \centering
  \begin{tabularx}{\textwidth}{|l|l|Y|Y|}
   \hline
   \textit{TD type} &  \textit{TD item} & \textit{TD dimension} & \textit{Identified By}  \\
   \hline
   Long method & Code smells & Code debt & DesigniteJava, Sonarcloud \\
   \hline
   Long parameter list & Code smells & Code debt & DesigniteJava, Sonarcloud \\
   \hline
   Complex method & Code smells & Code debt & DesigniteJava, Sonarcloud \\
   \hline
   Whitespace around & Coding guideline violation & Code debt & Checkstyle \\
   \hline
   Missing javadoc comment & Coding guideline violation & Code debt & Checkstyle \\ 
   \hline
   Add at least one assertion to this test case & Improper test design & Test debt & Sonarcloud \\
   \hline
   Coverage below $90\%$ & Inadequate test coverage & Test debt & Jacoco \\
   \hline
   Add some tests to this class & Lack of tests & Test debt & Sonarcloud, Jacoco \\
   \hline
   Deficient encapsulation & Design smells & Design debt & DesigniteJava \\
   \hline
   Hub-like modularization & Design smells & Design debt & DesigniteJava \\
   \hline
   Unutilized abstraction & Design smells & Design debt & DesigniteJava \\
   \hline
   Intercomponent cyclicality & Architecture smells & Architecture debt & Lattix \\
   \hline
\end{tabularx}
     \vspace{5pt}

    \caption{Some of the identified TD types, items, and dimensions from the chosen projects}
    \label{tab:identified}
\end{table}

However, tools like Lattix can identify a few more items like Architecture rule violations which fall under architecture debt. But for the tool to identify such violations, the rules should have been enabled at first. A glimpse of it is shown in the Appendix \ref{sec:demo}.

\subsection{Technical debt representation} \label{subsec:tdt}

The TD instances are mentioned in a structured tabular format in this subsection. The values in those tables were retrieved from different tools. For example, Codescene became handy to find the author responsible for a TD instance as it provides a rich social analysis.

\subsubsection{Java WebSocket} 

See tables \ref{tab:jws_start} to \ref{tab:jws_end}

\begin{table}[H]
    \begin{minipage}{0.49\textwidth}
         \begin{tabularx}{\textwidth}{|l|X|}
        \hline
        \textit{ID} & jws\_cd\_1\\
        \hline
        \textit{TD type name} & Long method\\
        \hline
        \textit{TD item name} & Code smells\\
        \hline
        \textit{Location} & Method decodeHandshake in class WebSocketImpl in package org.java\_websocket\\
        \hline
        \textit{Responsible/Author} & Davidiusdadi\\
        \hline
        \textit{Dimension} &  Code debt\\
        \hline
        \textit{Date/Time} & April 15, 2019\\
        \hline
        \textit{Context} & A private method in a Java concrete class.\\
        \hline
        \textit{Propagation} & Impacts other public methods in the same class that uses this method.\\
        \hline
        \textit{Intentionality} & Unintentional\\
        \hline
    \end{tabularx}
    \caption{}
    \label{tab:jws_start}
    \end{minipage}
    \begin{minipage}{0.49\textwidth}
         \begin{tabularx}{\textwidth}{|l|X|}
        \hline
        \textit{ID} & jws\_cd\_2\\
        \hline
        \textit{TD type name} & Whitespace around\\
        \hline
        \textit{TD item name} & Coding guideline violation\\
        \hline
        \textit{Location} & Line 193 in class AbstractWebSocket in package org.java\_websocket\\
        \hline
        \textit{Responsible/Author} & marci4\\
        \hline
        \textit{Dimension} &  Code debt \\
        \hline
        \textit{Date/Time} & April 15, 2019\\
        \hline
        \textit{Context} & A private method in a Java abstract class.\\
        \hline
        \textit{Propagation} & Impacts other public methods in the same and derived classes that makes use of this method. \\
        \hline
        \textit{Intentionality} & Unintentional\\
        \hline
    \end{tabularx}
    \caption{}
    \end{minipage}
 
\end{table}

\begin{table}[H]
    \begin{minipage}{0.49\textwidth}
         \begin{tabularx}{\textwidth}{|l|X|}
        \hline
        \textit{ID} & jws\_td\_1\\
        \hline
        \textit{TD type name} & Coverage below $90\%$ \\
        \hline
        \textit{TD item name} & Inadequate test coverage\\
        \hline
        \textit{Location} & All source files (coverage is only $64.2\%$)\\
        \hline
        \textit{Responsible/Author} & marci4, Marcel P\\
        \hline
        \textit{Dimension} & Test debt \\
        \hline
        \textit{Date/Time} & April 15, 2019\\
        \hline
        \textit{Context} & Jacoco coverage report for junit tests.\\
        \hline
        \textit{Propagation} & Impacts all source and test files.\\
        \hline
        \textit{Intentionality} & Intentional\\
        \hline
    \end{tabularx}
    \caption{}
    \end{minipage}
    \begin{minipage}{0.49\textwidth}
         \begin{tabularx}{\textwidth}{|l|X|}
        \hline
        \textit{ID} & jws\_td\_2\\
        \hline
        \textit{TD type name} & Add at least one assertion to this case\\
        \hline
        \textit{TD item name} & Improper test design\\
        \hline
        \textit{Location} & Line 151 in class Issue256Test in package org.java\_websocket.issues\\
        \hline
        \textit{Responsible/Author} & marci4\\
        \hline
        \textit{Dimension} & Test debt \\
        \hline
        \textit{Date/Time} & April 15, 2019\\
        \hline
        \textit{Context} & A junit test.\\
        \hline
        \textit{Propagation} & No impact to other classes.\\
        \hline
        \textit{Intentionality} & Unintentional\\
        \hline
    \end{tabularx}
    \caption{}
    \end{minipage}
 
\end{table}

\begin{table}[H]
    
    \begin{minipage}{0.49\textwidth}
         \begin{tabularx}{\textwidth}{|l|X|}
        \hline
        \textit{ID} & jws\_td\_3\\
        \hline
        \textit{TD type name} & Add some tests to this class \\
        \hline
        \textit{TD item name} & Lack of tests\\
        \hline
        \textit{Location} & Class AutobahnClientTest in package org.java\_websocket.example \\
        \hline
        \textit{Responsible/Author} & Davidiusdadi\\
        \hline
        \textit{Dimension} & Test debt\\
        \hline
        \textit{Date/Time} & April 15, 2019\\
        \hline
        \textit{Context} & A Java concrete class.\\
        \hline
        \textit{Propagation} & No impact to other classes.\\
        \hline
        \textit{Intentionality} & Intentional\\
        \hline
    \end{tabularx}
    \caption{}
    \end{minipage}
    \begin{minipage}{0.49\textwidth}
        \begin{tabularx}{\textwidth}{|l|X|}
        \hline
        \textit{ID} & jws\_dd\_1\\
        \hline
        \textit{TD type name} & Unutilized abstraction\\
        \hline
        \textit{TD item name} & Design smells\\
        \hline
        \textit{Location} & Class SSLSocketChannel in package org.java\_websocket\\
        \hline
        \textit{Responsible/Author} & marci4\\
        \hline
        \textit{Dimension} & Design debt\\
        \hline
        \textit{Date/Time} & April 15, 2019\\
        \hline
        \textit{Context} & A Java concrete class.\\
        \hline
        \textit{Propagation} & No impact to other classes.\\
        \hline
        \textit{Intentionality} & Unintentional\\
        \hline
        \end{tabularx}
    \caption{}
    \end{minipage}
 
\end{table}

\begin{table}[H]
    
    \begin{minipage}{0.49\textwidth}
         \begin{tabularx}{\textwidth}{|l|X|}
        \hline
        \textit{ID} & jws\_ad\_1\\
        \hline
        \textit{TD type name} & Intercomponent cyclicality ($9.67\%$)\\
        \hline
        \textit{TD item name} & Architecture smell\\
        \hline
        \textit{Location} & Classes in packages org.java\_websocket, org.java\_websocket.drafts and org.java\_websocket.server \\
        \hline
        \textit{Responsible/Author} &  marci4, Davidiusdadi\\
        \hline
        \textit{Dimension} & Architecture debt \\
        \hline
        \textit{Date/Time} & April 15, 2019\\
        \hline
        \textit{Context} & Java classes across different packages.\\
        \hline
        \textit{Propagation} & Impacts the classes that branches from the existing dependency cycles.\\
        \hline
        \textit{Intentionality} & Unintentional\\
        \hline
    \end{tabularx}
    \caption{}
    \label{tab:jws_end}
    \end{minipage}
 
\end{table}

\subsubsection{JDBM3}

See tables \ref{tab:jdb_start} to \ref{tab:jdb_end}.

\begin{table}[H]
    \begin{minipage}{0.49\textwidth}
         \begin{tabularx}{\textwidth}{|l|X|}
        \hline
        \textit{ID} & jdb\_cd\_1\\
        \hline
        \textit{TD type name} & Complex method\\
        \hline
        \textit{TD item name} & Code smells\\
        \hline
        \textit{Location} & Method equals in class SerialClassInfoTest in package org.apache.jdbm\\
        \hline
        \textit{Responsible/Author} & Jan Kotek\\
        \hline
        \textit{Dimension} &  Code debt\\
        \hline
        \textit{Date/Time} & April 15, 2019\\
        \hline
        \textit{Context} & A public method in a nested, static Java class.\\
        \hline
        \textit{Propagation} & Impacts other methods that calls this method.\\
        \hline
        \textit{Intentionality} & Unintentional\\
        \hline
    \end{tabularx}
    \caption{}
    \label{tab:jdb_start}
    \end{minipage}
    \begin{minipage}{0.49\textwidth}
         \begin{tabularx}{\textwidth}{|l|X|}
        \hline
        \textit{ID} & jdb\_cd\_2\\
        \hline
        \textit{TD type name} & Switch without \enquote{default} clause\\
        \hline
        \textit{TD item name} & Coding guideline violation\\
        \hline
        \textit{Location} & Line 633 in class Serialization in package org.apache.jdbm\\
        \hline
        \textit{Responsible/Author} & Jan Kotek\\
        \hline
        \textit{Dimension} &  Code debt\\
        \hline
        \textit{Date/Time} & April 15, 2019\\
        \hline
        \textit{Context} & A public method in a public Java concrete class.\\
        \hline
        \textit{Propagation} & Impacts other methods that calls this method.\\
        \hline
        \textit{Intentionality} & Unintentional \\
        \hline
    \end{tabularx}
    \caption{}
    \end{minipage}
 
\end{table}

\begin{table}[H]
    \begin{minipage}{0.49\textwidth}
         \begin{tabularx}{\textwidth}{|l|X|}
        \hline
        \textit{ID} & jdb\_td\_1\\
        \hline
        \textit{TD type name} & Coverage below $90\%$ \\
        \hline
        \textit{TD item name} & Inadequate test coverage\\
        \hline
        \textit{Location} & All source files (coverage is only $82.6\%$)\\
        \hline
        \textit{Responsible/Author} & Jan Kotek\\
        \hline
        \textit{Dimension} & Test debt\\
        \hline
        \textit{Date/Time} & April 15, 2019\\
        \hline
        \textit{Context} & Jacoco coverage report for junit tests.\\
        \hline
        \textit{Propagation} & Impacts all source and test files.\\
        \hline
        \textit{Intentionality} & Unintentional\\
        \hline
    \end{tabularx}
    \caption{}
    \end{minipage}
    \begin{minipage}{0.49\textwidth}
         \begin{tabularx}{\textwidth}{|l|X|}
        \hline
        \textit{ID} & jdb\_td\_2\\
        \hline
        \textit{TD type name} & Add at least one assertion to this case\\
        \hline
        \textit{TD item name} & Improper test design\\
        \hline
        \textit{Location} & Line 91 in class BTreeTest in package org.apache.jdbm\\
        \hline
        \textit{Responsible/Author} & Jan Kotek\\
        \hline
        \textit{Dimension} & Test debt \\
        \hline
        \textit{Date/Time} & April 15, 2019\\
        \hline
        \textit{Context} & A junit test in a Java class.\\
        \hline
        \textit{Propagation} & No impact to other classes. \\
        \hline
        \textit{Intentionality} & Unintentional\\
        \hline
    \end{tabularx}
    \caption{}
    \end{minipage}
 
\end{table}

\begin{table}[H]

    \begin{minipage}{0.49\textwidth}
        \begin{tabularx}{\textwidth}{|l|X|}
        \hline
        \textit{ID} & jdb\_dd\_1\\
        \hline
        \textit{TD type name} & Deficient encapsulation\\
        \hline
        \textit{TD item name} & Design smells\\
        \hline
        \textit{Location} & Class DBStore in package org.apache.jdbm\\
        \hline
        \textit{Responsible/Author} & Jan Kotek\\
        \hline
        \textit{Dimension} & Design debt\\
        \hline
        \textit{Date/Time} & April 15, 2019\\
        \hline
        \textit{Context} & A Java concrete class.\\
        \hline
        \textit{Propagation} & Impacts methods that calls or makes use of that method or attribute.\\
        \hline
        \textit{Intentionality} & Intentional\\
        \hline
        \end{tabularx}
    \caption{}
    \label{tab:jdb_end}
    \end{minipage}
\end{table}

\subsubsection{Jedis}

See tables \ref{tab:jed_start} to \ref{tab:jed_end}

\begin{table}[H]
    \begin{minipage}{0.49\textwidth}
         \begin{tabularx}{\textwidth}{|l|X|}
        \hline
        \textit{ID} & jed\_cd\_1\\
        \hline
        \textit{TD type name} & Long parameter list\\
        \hline
        \textit{TD item name} & Code smells\\
        \hline
        \textit{Location} & Constructor in class JedisClusterInfoCache in package redis.clients.jedis\\
        \hline
        \textit{Responsible/Author} & Jungtaek Lim\\
        \hline
        \textit{Dimension} &  Code debt\\
        \hline
        \textit{Date/Time} & April 15, 2019\\
        \hline
        \textit{Context} & A public method in a Java concrete class.\\
        \hline
        \textit{Propagation} & Impacts methods that calls this method.\\
        \hline
        \textit{Intentionality} & Unintentional\\
        \hline
    \end{tabularx}
    \caption{}
    \label{tab:jed_start}
    \end{minipage}
    \begin{minipage}{0.49\textwidth}
         \begin{tabularx}{\textwidth}{|l|X|}
        \hline
        \textit{ID} & jed\_cd\_2\\
        \hline
        \textit{TD type name} & Unicode escapes should be avoided\\
        \hline
        \textit{TD item name} & Coding guideline violation\\
        \hline
        \textit{Location} & Line 161 in class BitCommandsTest in package redis.clients.jedis.tests.commands\\
        \hline
        \textit{Responsible/Author} & Marcos Nils \\
        \hline
        \textit{Dimension} &  Code debt\\
        \hline
        \textit{Date/Time} & April 15, 2019\\
        \hline
        \textit{Context} & A junit test in a Java class.\\
        \hline
        \textit{Propagation} & No impact to other classes.\\
        \hline
        \textit{Intentionality} & Intentional\\
        \hline
    \end{tabularx}
    \caption{}
    \end{minipage}
 
\end{table}

\begin{table}[H]
    \begin{minipage}{0.49\textwidth}
         \begin{tabularx}{\textwidth}{|l|X|}
        \hline
        \textit{ID} & jed\_td\_1\\
        \hline
        \textit{TD type name} & Coverage below $90\%$ \\
        \hline
        \textit{TD item name} & Inadequate test coverage\\
        \hline
        \textit{Location} & All source files (coverage is only $10.7\%$)\\
        \hline
        \textit{Responsible/Author} & Jonathan Leibiusky\\
        \hline
        \textit{Dimension} & Test debt\\
        \hline
        \textit{Date/Time} & April 15, 2019\\
        \hline
        \textit{Context} & Jacoco coverage report for junit tests.\\
        \hline
        \textit{Propagation} & Impact to all source and test files. \\
        \hline
        \textit{Intentionality} & Intentional \\
        \hline
    \end{tabularx}
    \caption{}
    \end{minipage}
    \begin{minipage}{0.49\textwidth}
         \begin{tabularx}{\textwidth}{|l|X|}
        \hline
        \textit{ID} & jed\_td\_2\\
        \hline
        \textit{TD type name} & Add at least one assertion to this case\\
        \hline
        \textit{TD item name} & Improper test design\\
        \hline
        \textit{Location} & Line 111 in class SSLJedisClusterTest in package redis.clients.jedis.tests\\
        \hline
        \textit{Responsible/Author} & M Sazzadul Hoque\\
        \hline
        \textit{Dimension} & Test debt \\
        \hline
        \textit{Date/Time} & April 15, 2019\\
        \hline
        \textit{Context} & A junit test in a Java class.\\
        \hline
        \textit{Propagation} & No impact to other classes.\\
        \hline
        \textit{Intentionality} & Unintentional\\
        \hline
    \end{tabularx}
    \caption{}
    \end{minipage}
 
\end{table}

\begin{table}[H]

    \begin{minipage}{0.49\textwidth}
        \begin{tabularx}{\textwidth}{|l|X|}
        \hline
        \textit{ID} & jed\_dd\_1\\
        \hline
        \textit{TD type name} & Insufficient modularization\\
        \hline
        \textit{TD item name} & Design smells\\
        \hline
        \textit{Location} & Interface JedisClusterCommands in package redis.clients.jedis.commands\\
        \hline
        \textit{Responsible/Author} & phufool\\
        \hline
        \textit{Dimension} & Design debt \\
        \hline
        \textit{Date/Time} & April 15, 2019\\
        \hline
        \textit{Context} & A Java interface.\\
        \hline
        \textit{Propagation} & Impacts all the clients of this interface.\\
        \hline
        \textit{Intentionality} & Intentional\\
        \hline
        \end{tabularx}
    \caption{}
    \label{tab:jed_end}
    \end{minipage}
 
\end{table}

\subsubsection{MyBatis}

See tables \ref{tab:myb_start} to \ref{tab:myb_end}.

\begin{table}[H]
    \begin{minipage}{0.49\textwidth}
         \begin{tabularx}{\textwidth}{|l|X|}
        \hline
        \textit{ID} & myb\_cd\_1\\
        \hline
        \textit{TD type name} & Magic number\\
        \hline
        \textit{TD item name} & Code smells\\
        \hline
        \textit{Location} & Line 77 in class CacheTest in package org.apache.ibattis.submitted.cache\\
        \hline
        \textit{Responsible/Author} & Kazuki Shimizu\\
        \hline
        \textit{Dimension} &  Code debt\\
        \hline
        \textit{Date/Time} & April 15, 2019\\
        \hline
        \textit{Context} & A juni test in a Java class.\\
        \hline
        \textit{Propagation} & No impact to other classes.\\
        \hline
        \textit{Intentionality} & Unintentional\\
        \hline
    \end{tabularx}
    \caption{}
    \label{tab:myb_start}
    \end{minipage}
    \begin{minipage}{0.49\textwidth}
         \begin{tabularx}{\textwidth}{|l|X|}
        \hline
        \textit{ID} & myb\_cd\_2\\
        \hline
        \textit{TD type name} & Line is longer than 100 characters\\
        \hline
        \textit{TD item name} & Coding guideline violation\\
        \hline
        \textit{Location} & Line 47 in class DeleteProvider in package org.apache.ibatis.annotations\\
        \hline
        \textit{Responsible/Author} & Kazuki Shimizu\\
        \hline
        \textit{Dimension} &  Code debt \\
        \hline
        \textit{Date/Time} & April 15, 2019\\
        \hline
        \textit{Context} & Docstring for a Java interface.\\
        \hline
        \textit{Propagation} & No impact to other classes.\\
        \hline
        \textit{Intentionality} & Unintentional\\
        \hline
    \end{tabularx}
    \caption{}
    \end{minipage}
 
\end{table}

\begin{table}[H]
    \begin{minipage}{0.49\textwidth}
         \begin{tabularx}{\textwidth}{|l|X|}
        \hline
        \textit{ID} & myb\_td\_1\\
        \hline
        \textit{TD type name} & Coverage below $90\%$ \\
        \hline
        \textit{TD item name} & Inadequate test coverage\\
        \hline
        \textit{Location} & All source files (coverage is only $84.3\%$)\\
        \hline
        \textit{Responsible/Author} & Eduardo Macarron, Nathan Maves, Iwao Ave
        \\
        \hline
        \textit{Dimension} & Test debt\\
        \hline
        \textit{Date/Time} & April \\
        \hline
        \textit{Context} & Jacoco coverage report for junit tests.\\
        \hline
        \textit{Propagation} & Impacts all source and test files.\\
        \hline
        \textit{Intentionality} & Unintentional\\
        \hline
    \end{tabularx}
    \caption{}
    \end{minipage}
    \begin{minipage}{0.49\textwidth}
         \begin{tabularx}{\textwidth}{|l|X|}
        \hline
        \textit{ID} & myb\_td\_2\\
        \hline
        \textit{TD type name} & Add at least one assertion to this case\\
        \hline
        \textit{TD item name} & Improper test design\\
        \hline
        \textit{Location} & Line 390 in class BindingTest in package org.apache.ibatis.binding\\
        \hline
        \textit{Responsible/Author} & Eduardo Macarron\\
        \hline
        \textit{Dimension} & Test debt \\
        \hline
        \textit{Date/Time} & April 15, 2019\\
        \hline
        \textit{Context} & A junit test in a Java class.\\
        \hline
        \textit{Propagation} & No impact to other classes.\\
        \hline
        \textit{Intentionality} & Unintentional\\
        \hline
    \end{tabularx}
    \caption{}
    \end{minipage}
 
\end{table}

\begin{table}[H]

    \begin{minipage}{0.49\textwidth}
        \begin{tabularx}{\textwidth}{|l|X|}
        \hline
        \textit{ID} & myb\_dd\_1\\
        \hline
        \textit{TD type name} & Unnecessary abstraction\\
        \hline
        \textit{TD item name} & Design smells\\
        \hline
        \textit{Location} & Class StaticClass in package org.apache.ibatis.submitted.ognlstatic\\
        \hline
        \textit{Responsible/Author} & Eduardo Macarron \\
        \hline
        \textit{Dimension} & Design debt \\
        \hline
        \textit{Date/Time} & April 15, 2019\\
        \hline
        \textit{Context} & A Java static class.\\
        \hline
        \textit{Propagation} & Impacts classes that use this class.\\
        \hline
        \textit{Intentionality} & Intentional\\
        \hline
        \end{tabularx}
    \caption{}
    \end{minipage}
     \begin{minipage}{0.49\textwidth}
         \begin{tabularx}{\textwidth}{|l|X|}
        \hline
        \textit{ID} & myb\_ad\_1\\
        \hline
        \textit{TD type name} & Intercomponent cyclicality ($11.655\%$)\\
        \hline
        \textit{TD item name} & Architecture smell\\
        \hline
        \textit{Location} & Classes in annotations, binding, builder, executor, mapping and a few more packages within namespace org.apache.batis\\
        \hline
        \textit{Responsible/Author} & Nathan Maves\\
        \hline
        \textit{Dimension} & Architecture debt \\
        \hline
        \textit{Date/Time} & April 15, 2019\\
        \hline
        \textit{Context} & Java classes across different packages.\\
        \hline
        \textit{Propagation} &  Impacts the classes that branches from the existing dependency cycles.\\
        \hline
        \textit{Intentionality} & Unintentional\\
        \hline
    \end{tabularx}
    \caption{}
    \label{tab:myb_end}
    \end{minipage}
 
\end{table}

\begin{table}[H]

\end{table}

\subsection{Technical debt estimation}

Out of all the used tools, only Sonarcloud gives an estimate of TD principal by default. However, the estimate doesn't take into account all of the TD items across all the TD dimensions. The estimate is calculated only based on the following TD items: code smells, lack of tests, and improper test design.

\begin{table}[h!]
  \centering
    \begin{tabular}{|c|c|}
       \hline
       \textit{Project} & \textit{TD principal estimate} \\
       \hline
       Java WebSocket & 3 \\
       \hline
       JDBM3 &  9 \\
       \hline
       Jedis & 10\\
       \hline
       MyBatis & 20\\
       \hline
    \end{tabular}
    \caption{TD principal estimates from Sonarcloud in terms of person-days}
    \label{tab:td_principal_estimates}
\end{table}

\subsection{Technical debt monitoring} \label{subsec:monitoring}

\begin{figure}[H]
\centering
\includegraphics[width=\columnwidth,height=0.45\textheight,keepaspectratio]{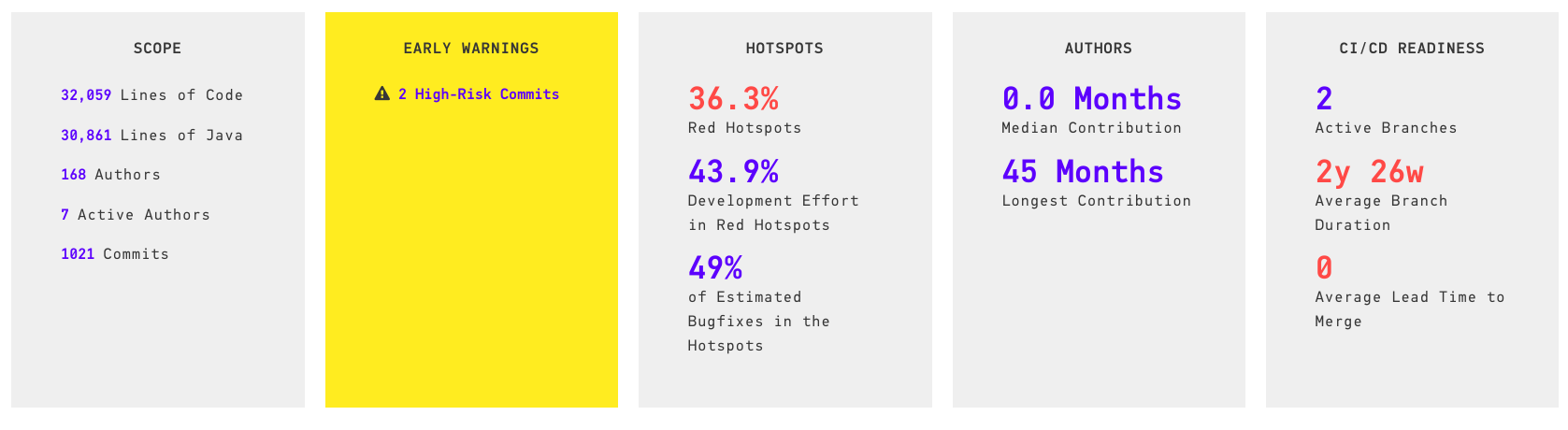}
\captionof{figure}{Dashboard from Codescene for one of the project \enquote{Jedis}}
\label{fig:cs_db}
\end{figure}

\begin{figure}[H]
\centering
\includegraphics[width=\columnwidth,height=0.45\textheight,keepaspectratio]{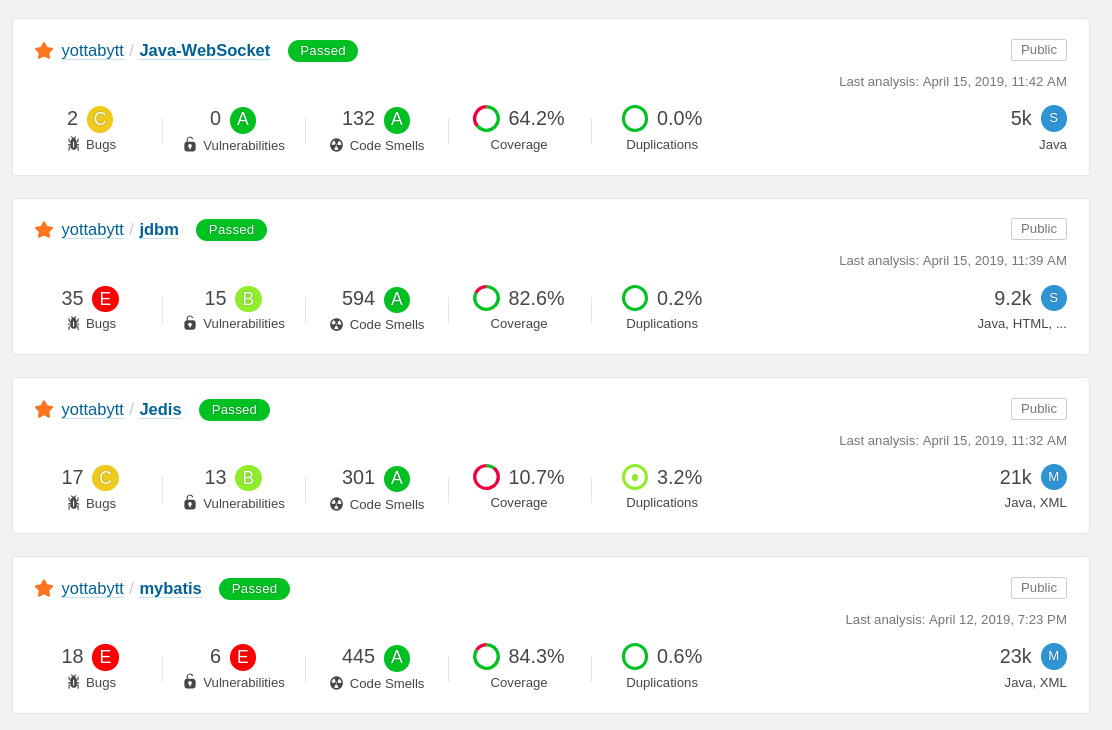}
\captionof{figure}{Dashboard from Sonarcloud for all of the chosen projects}
\label{fig:sc_db}
\end{figure}

Dashboards [See figures \ref{fig:sc_db} and \ref{fig:cs_db}] and warnings/alerts [See figure \ref{fig:cs_db}] can be enabled by integrating some of the tools with the IDE or with the continuous integration (CI) servers. Let's say Sonarqube is integrated with Jenkins, then developers and product owners can be kept informed about the TD instances that has happened or might soon happen because of the recent commits. Tools like Codescene  can directly look for the commits that is been made to a repository and can warn the stakeholders by re-running the analysis and producing the reports.

\subsection{Technical debt repayment}

In this subsection, we propose techniques to repay the principal of three TD instances for every chosen project. 

\begin{table}[H]

    \begin{minipage}{0.49\textwidth}
        \begin{tabularx}{\textwidth}{|l|X|}
        \hline
        \textit{ID} & \textit{Proposed techniques for repayment}\\
        \hline
        jws\_cd\_1 & Split the method decodeHandshake into multiple methods by extracting code from the branch statements i.e., make the body of branch statements as individual methods.\\
        \hline
        jws\_cd\_2 & Add whitespace around all the symbols in line 193 as per Google style guide for Java. \\
        \hline
        jws\_ad\_1 & Many classes in org.java\_websocket depend on classes from other packages. To reduce the percent of intercomponent cyclicality, either move the coupled classes into same package if possible or introduce a bridge class in current package and make it to talk to classes in other packages. \\
        \hline
        \end{tabularx}
    \caption{TD repayment for Java WebSocket}
    \end{minipage}
     \begin{minipage}{0.49\textwidth}
         \begin{tabularx}{\textwidth}{|l|X|}
        \hline
        \textit{ID} & \textit{Proposed techniques for repayment}\\
        \hline
        jdb\_cd\_1 & Move some of the conditional statements into a separate method or try making use of polymorphism. \\
       \hline
       jdb\_cd\_2 & Add a default case in the switch block.\\
       \hline
       jdb\_dd\_1 & In line 115, change the public modifier to private or protected. \\
       \hline
    \end{tabularx}
    \caption{TD repayment for JDBM3}
    \end{minipage}
 
\end{table}

\begin{table}[H]

    \begin{minipage}{0.49\textwidth}
        \begin{tabularx}{\textwidth}{|l|X|}
        \hline
        \textit{ID} & \textit{Proposed techniques for repayment}\\
        \hline
        jed\_cd\_1 & Group the parameters into some collection data structure.\\
        \hline
        jed\_td\_2 & Add an assertion statement either by comparing to connection status or to the value retrieved. \\
        \hline
        jed\_dd\_1 & The interface seem to have lot of methods. It can be broken down into many interfaces by grouping similar client-specific methods together. \\
        \hline
        \end{tabularx}
    \caption{TD repayment for Jedis}
    \end{minipage}
     \begin{minipage}{0.49\textwidth}
         \begin{tabularx}{\textwidth}{|l|X|}
        \hline
        \textit{ID} & \textit{Proposed techniques for repayment}\\
        \hline
        myb\_cd\_1 & Introduce a variable and assign it to an integer with value 1 and use that variable instead. \\
        \hline
        myb\_cd\_2 & Break the long line within $<$li$>$ tag by introducing a $<$br$>$ tag to keep the number of characters less than $100$.\\
        \hline
        myb\_td\_1 & Add more tests to source files which are far below the set threshold till the project coverage crosses the threshold.\\
       \hline
    \end{tabularx}
    \caption{TD repayment for MyBatis}
    \end{minipage}
 
\end{table}

\begin{figure}[H]
  \centering
  \includegraphics[width=\columnwidth,height=0.45\textheight,keepaspectratio]{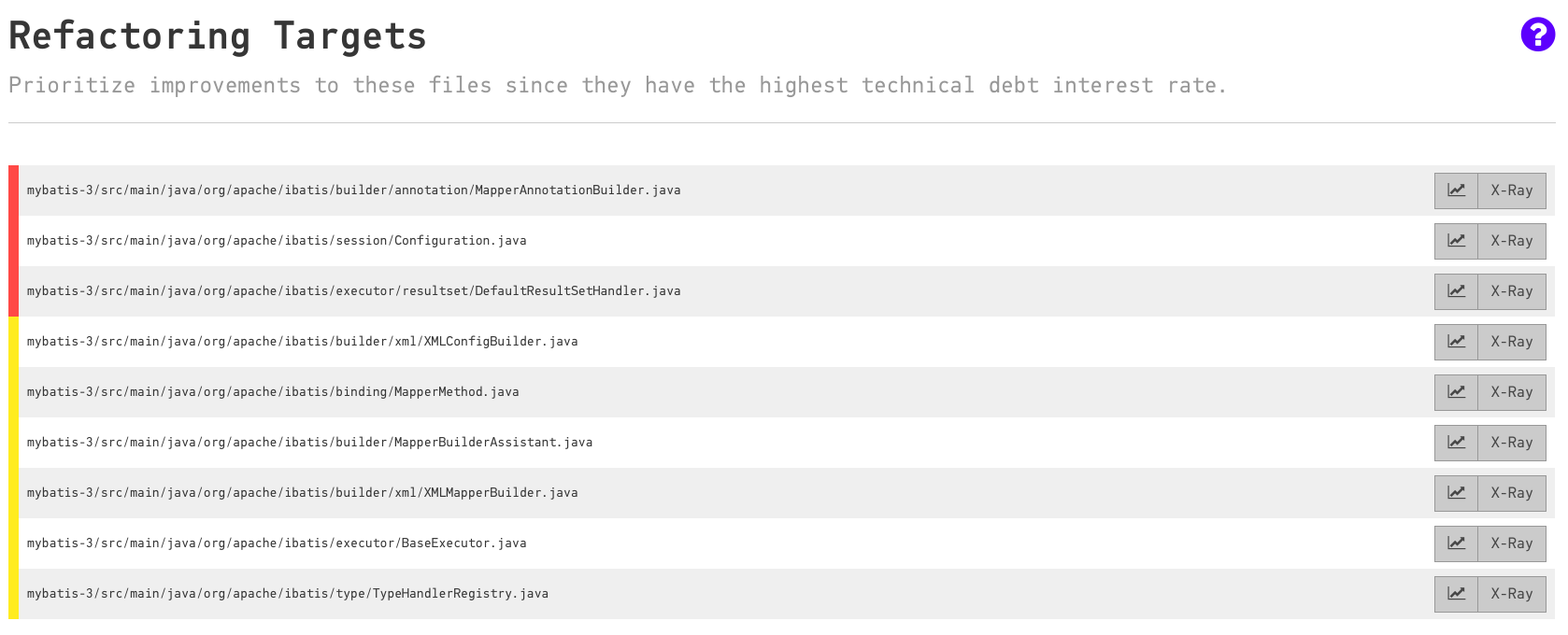}
  \captionof{figure}{Refactoring targets for MyBatis}
  \label{fig:cs_refact}
\end{figure}

However, in reality, acting upon immediately on all the TD instances is not worthy. There are tools like Codescene which helps in prioritizing the refactoring targets. It prioritizes TD instances based on their technical debt interest rate. Look at the screenshot [figure \ref{fig:cs_refact}] from Codescene for one of the projects.

\subsection{Technical debt prevention}

There are no tools out there than can automatically prevent the occurence of a TD. Because, it happens mostly due to human choices and mistakes. However, with rich information that could be exposed from the source code repositories [See \ref{subsec:monitoring}], we can prevent a TD instance from getting deployed into production systems. Also, once the developers get to know their mistakes with the help of such tools, the frequency of the same TD type getting introduced in the future can gradually get decreased.

\subsection{Challenges faced}

It was never a straightforward process of selecting the projects, feeding them into multiple tools, and observing the results. We overcame several limitations to present our empirical observations in a coherent manner. Here is a glimpse on some of the challenges which were worth mentioning, 
\begin{itemize}
\item As many tools were very much similar to each other in terms of their functionality, choosing a diverse set of tools to cover many TD dimensions was the first and foremost challenge. Few tools were not free. So it took a couple of email conversations to get a limited time access. 
\item DesigniteJava quickly runs out of memory on a $12$GB machine for projects involving $>100K$ LOC. As we wanted results for a chosen project from all the chosen tools, we had to choose projects which were not huge yet not small.
\item For few tools, integration with Maven and Gradle were not consistent as both of those build tools behave differently. So we decided to stick just to Maven projects. But searching for Maven \href{https://github.com/search?l=Java\&type=Repositories 
}{Java} projects in GitHub was slightly time consuming as it seemed to have been outnumbered by Gradle projects.
\item Another factor in the abovementioned time consuming search process was Jacoco. It's out-of-the box support for multi-module Maven projects was not simple. Because of that, for such projects, coverage was reported as $0\%$ even when they had tests. So we had to limit ourselves to single-module Maven projects. But we are confident that by spending a little more time on configurations and customizations, multi-module projects can be made to work.
\end{itemize}

\section{Proposal}

\subsection{Cost model}
Here we propose a simple cost model for estimating the TD prinicipal similar to the one present in Sonarqube TD plugin. But here we consider only till the level of TD item and not the TD type [See \ref{tab:identified}].  
\def\arraystretch{1.4}%
\begin{table}[H]
	\centering
	\begin{tabularx}{0.9\columnwidth}{|l|Y|}
	\hline
	\textit{Cost} & \textit{Default value (in person-hours)} \\
	\hline
	cost\_to\_fix\_a\_code\_smell & 5\\
	\hline
	cost\_to\_fix\_a\_coding\_guideline\_violation & 1\\
	\hline
	cost\_to\_fix\_an\_improper\_test\_design & 4\\
	\hline
	cost\_to\_fix\_a\_lack\_of\_test & 2\\
	\hline
	cost\_to\_fix\_inadequate\_test\_coverage (project level) & (difference between set threshold and current coverage) $\times$ $\frac{\text{Project's LOC}}{1000}$ \\
	\hline
	cost\_to\_fix\_a\_design\_smell & 15\\
	\hline
	cost\_to\_fix\_an\_architecture\_smell & 25\\
	\hline
	\end{tabularx}
	\caption{A cost model for estimating TD principal}
	\label{tab:costmodel}
\end{table}
\def\arraystretch{1.2}%
So, the general simple formula would be, 
\begin{align*}
\text{TD principal estimate} &= \text{cost\_to\_fix\_a\_code\_smell} \times \#\{\text{code smells}\} \\
& + \text{cost\_to\_fix\_a\_coding\_guideline\_violation} \times \#\{\text{coding guideline violations}\} \\
& + \text{cost\_to\_fix\_an\_improper\_test\_design}
\times \#\{\text{improper test designs}\} \\
& + \text{cost\_to\_fix\_a\_lack\_of\_test} 
\times \#\{\text{lack of tests}\} \\
& + (\text{expected coverage} - \text{current coverage})
\times \frac{\text{Project's LOC}}{1000}\\
& + \text{cost\_to\_fix\_a\_design\_smell} \times \#\{\text{design smells}\}\\
& + \text{cost\_to\_fix\_an\_architecture\_smell} \times \#\{\text{architecture smells}\}\\
\end{align*}

Now, with the cost model as mentioned in the above table \ref{tab:costmodel}, we estimate the TD principal for the chosen projects but by only considering the instances that were represented as multiple tables within the subsection \ref{subsec:tdt}. \\

\begin{itemize}
\item Java WebSocket \par
\begin{align*}
\text{TD principal estimate} &= 5 \times 1 (\#\{\text{jws\_cd\_1}\}) + 1 \times 1 (\#\{\text{jws\_cd\_2}\}) + 4 \times 1 (\#\{\text{jws\_td\_2}\}) + 2 \times 1 (\#\{\text{jws\_td\_3}\}) \\
& + (90 - 64.2) \times \frac{5000}{1000} + 15 \times 1 (\#\{\text{jws\_dd\_1}\}) + 25 \times 1 (\#\{\text{jws\_ad\_1}\}) \\
& = 5 + 1 + 4 + 2 + 129 + 15 + 25 \\
& = 181 \text{ person-hours}
\end{align*}
\item JDBM3 \par
\begin{align*}
\text{TD principal estimate} &= 5 \times 1 (\#\{\text{jdb\_cd\_1}\}) + 1 \times 1 (\#\{\text{jdb\_cd\_2}\}) + 4 \times 1 (\#\{\text{jdb\_td\_2}\})  \\
& + (90 - 82.6) \times \frac{9100}{1000} + 15 \times 1 (\#\{\text{jdb\_dd\_1}\}) \\
& = 5 + 1 + 4 + 67.34 + 15 \\
& = 92.34 \text{ person-hours}
\end{align*}
\item Jedis \par
\begin{align*}
\text{TD principal estimate} &= 5 \times 1 (\#\{\text{jed\_cd\_1}\}) + 1 \times 1 (\#\{\text{jed\_cd\-2}\}) + 4 \times 1 (\#\{\text{jed\_td\_2}\}) \\
& + (90 - 10.7) \times \frac{20800}{1000} + 15 \times 1 (\#\{\text{jed\_dd\_1}\}) \\
& = 5 + 1 + 4 + 1649.44 + 15 \\
& = 1674.44 \text{ person-hours}
\end{align*}
\item MyBatis \par
\begin{align*}
\text{TD principal estimate} &= 5 \times 1 (\#\{\text{myb\_cd\_1}\}) + 1 \times 1 (\#\{\text{myb\_cd\_2}\}) + 4 \times 1 (\#\{\text{myb\_td\_2}\}) \\
& + (90 - 84.3) \times \frac{22600}{1000} + 15 \times 1 (\#\{\text{myb\_dd\_1}\}) + 25 \times 1 (\#\{\text{myb\_ad\_1}\}) \\
& = 178.82 \text{ person-hours}
\end{align*}
\end{itemize}

\subsection{More tools to manage TD}

\href{https://deepsource.io/}{DeepSource}, a tool which is relatively new and got released for users during the month of November 2018, is something practitioners should keep an eye on. The team behind it seem to rapidly expand the feature set and support for multiple languages. The important thing is that the tool has a very neat and an elegant UI, a \href{https://deepsource.io/docs/}{clear documentation} of what it has to offer, and a responsive support. Also, to run the initial analysis, DeepSource is similar to Codacy, Codescene and dissimilar to Sonarcloud (without a Continuous Integration setup). It is integrated directly to the GitHub accounts and runs the initial analysis by cloning the repositories directly to their servers. Below is a screenshot [figure \ref{fig:deepsource}] from its website that mentions about some of their fully-available and preview features,

\begin{figure}[H]
\centering
\includegraphics[width=\columnwidth,height=0.45\textheight,keepaspectratio]{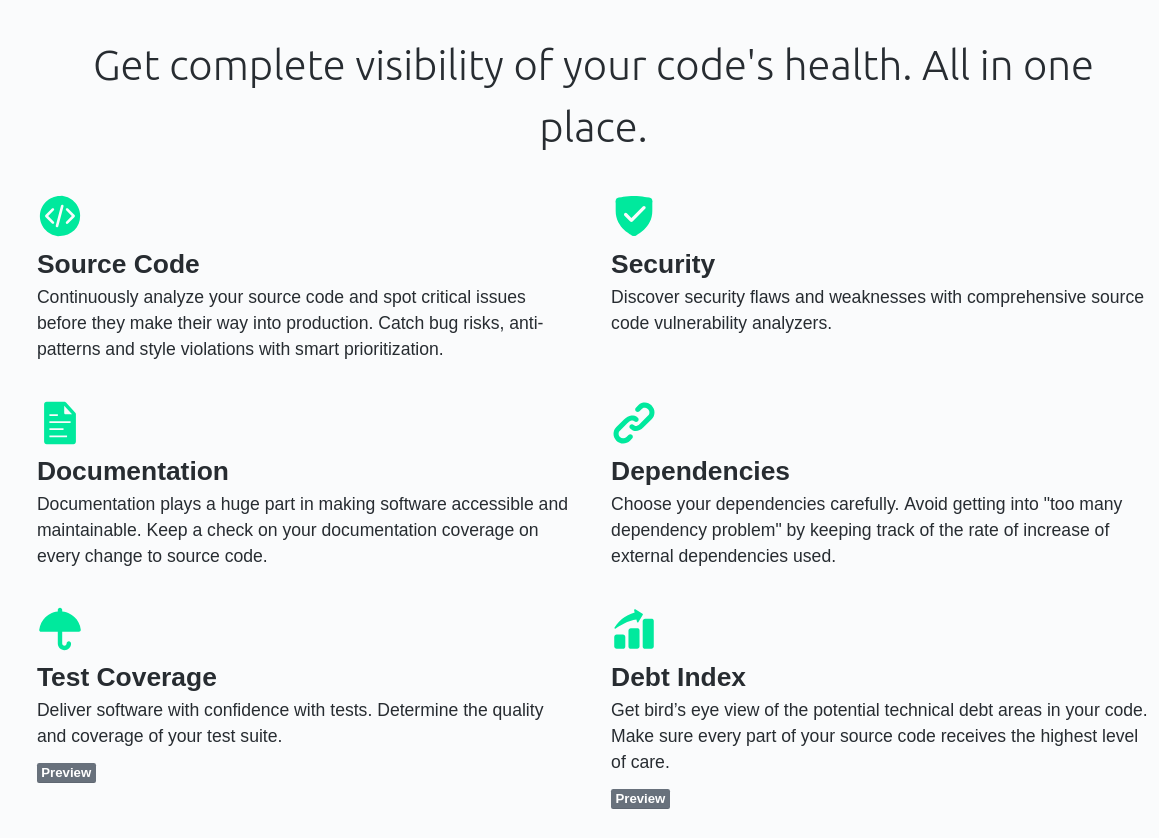}
\captionof{figure}{Some of the fully-available and preview features from DeepSource}
\label{fig:deepsource}
\end{figure}

Looking at figure \ref{fig:deepsource}, there are two out-of-the box feature that easily makes DeepSource standout among its peers. Firstly, its ability to address \enquote{Documentation Debt}. Secondly, the tool's ability to find issues with dependencies which quickly reminds us of the \enquote{Build Dependency Debt} introduced by Google \cite{build_debt}.

\begin{figure}[H]
\centering
\includegraphics[width=\columnwidth,height=0.45\textheight,keepaspectratio]{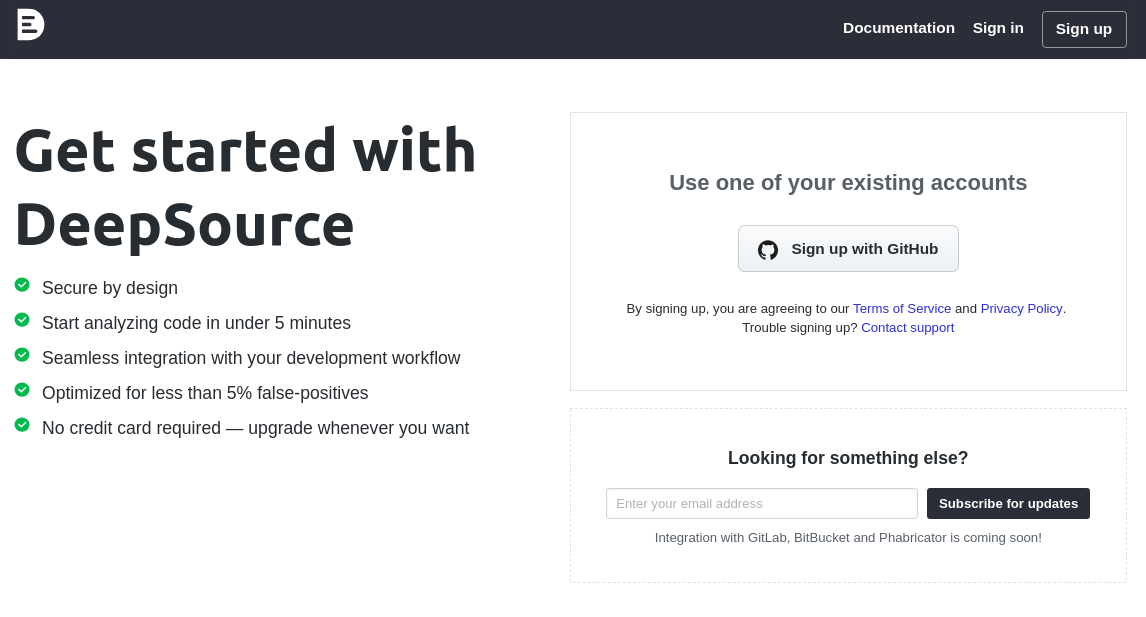}
\captionof{figure}{DeepSource - \href{https://deepsource.io/signup/}{Sign up page}. Click on Sign up with GitHub}
\label{fig:ds_start}
\end{figure}

\begin{figure}[H]
\centering
\includegraphics[width=\columnwidth,height=0.45\textheight,keepaspectratio]{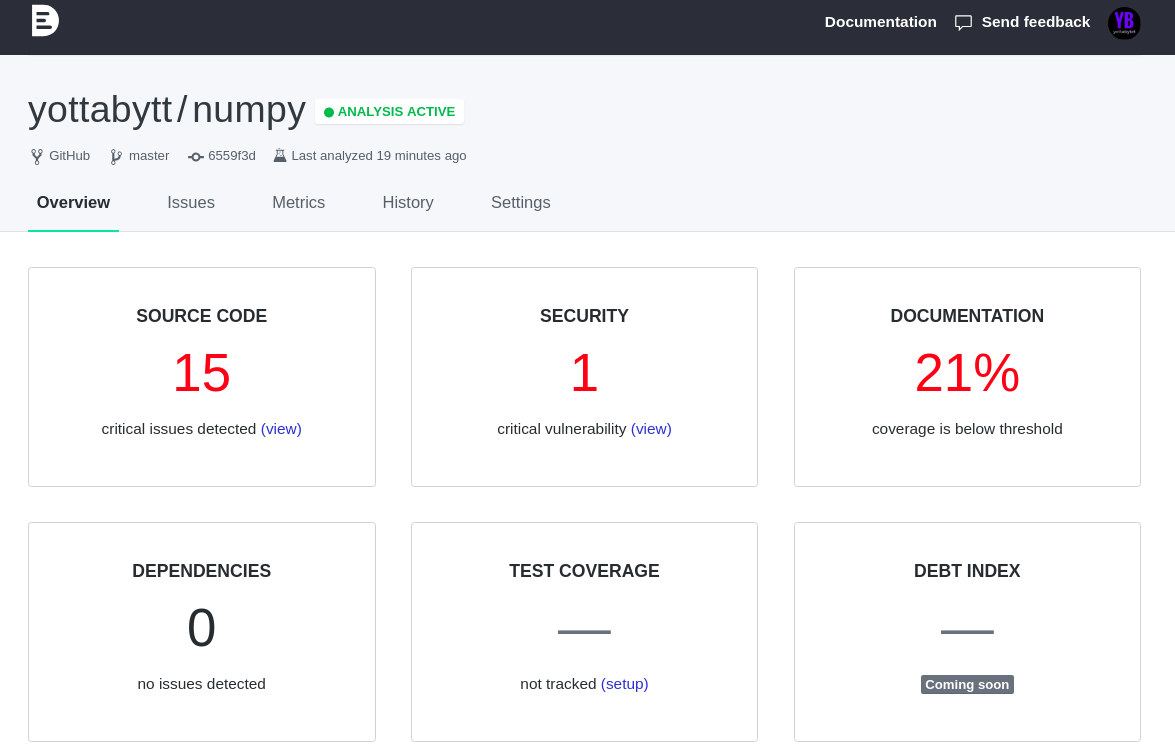}
\captionof{figure}{DeepSource - Overview tab that gives high-level information about the analysis.}
\label{fig:ds_end}
\end{figure}

Figures \ref{fig:ds_start} to \ref{fig:ds_end} gives a walk-through of the steps involved(in DeepSource) to run an analysis on a source code repository. The captions of those figures aid the screenshots with a description of what next to do.

\begin{figure}[H]
\centering
\includegraphics[width=\columnwidth,height=0.45\textheight,keepaspectratio]{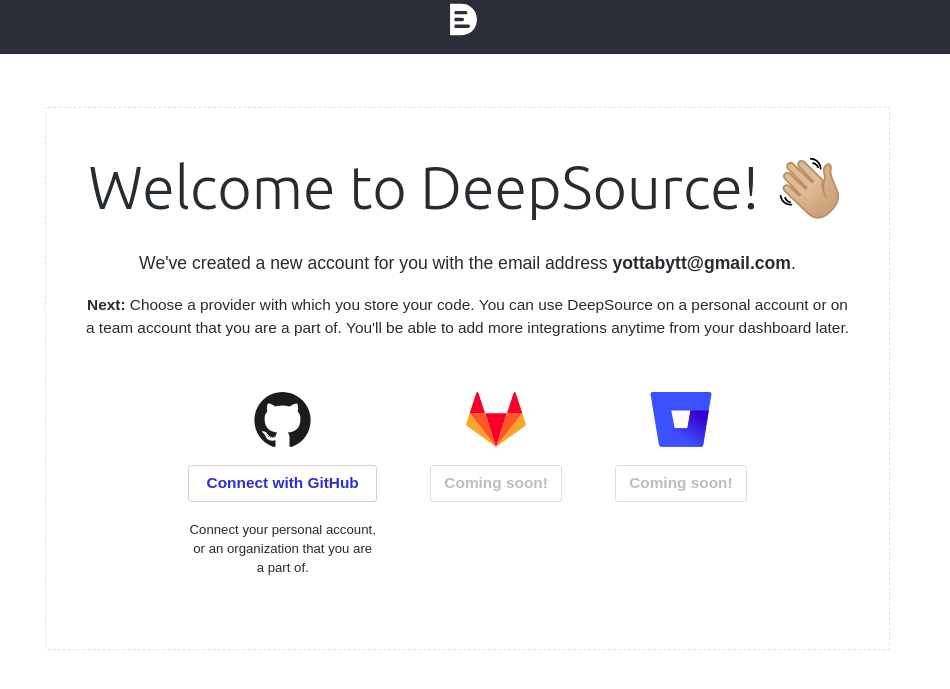}
\captionof{figure}{DeepSource - Sign up process.}
\label{fig:deepsource}
\end{figure}

\begin{figure}[H]
\centering
\includegraphics[width=\columnwidth,height=0.45\textheight,keepaspectratio]{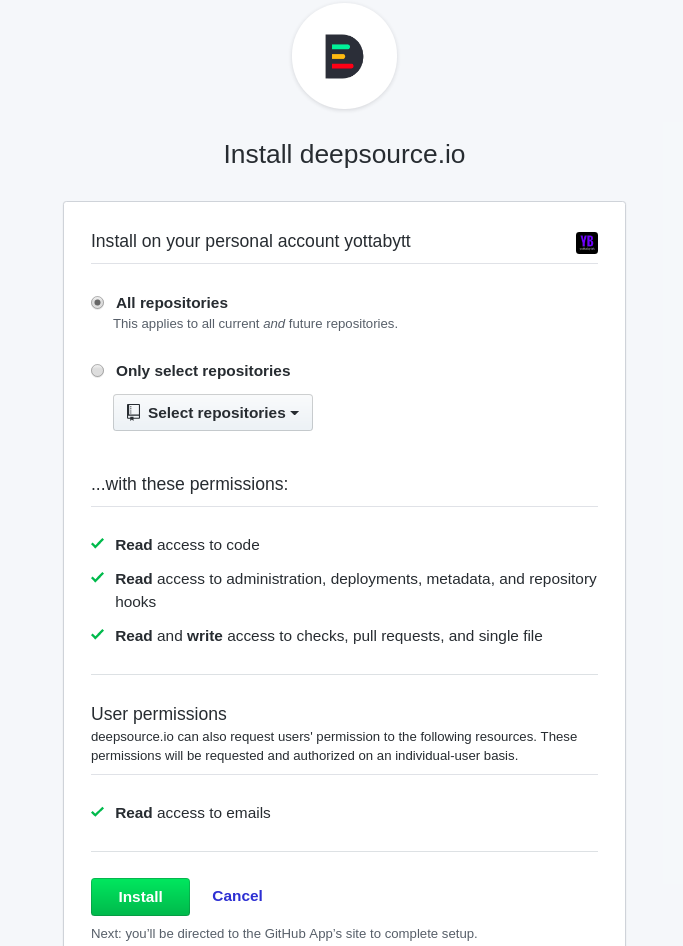}
\captionof{figure}{DeepSource - Sign up process. Grant appropriate permissions to the tool before installing.}
\label{fig:deepsource}
\end{figure}

\begin{figure}[H]
\centering
\includegraphics[width=\columnwidth,height=0.45\textheight,keepaspectratio]{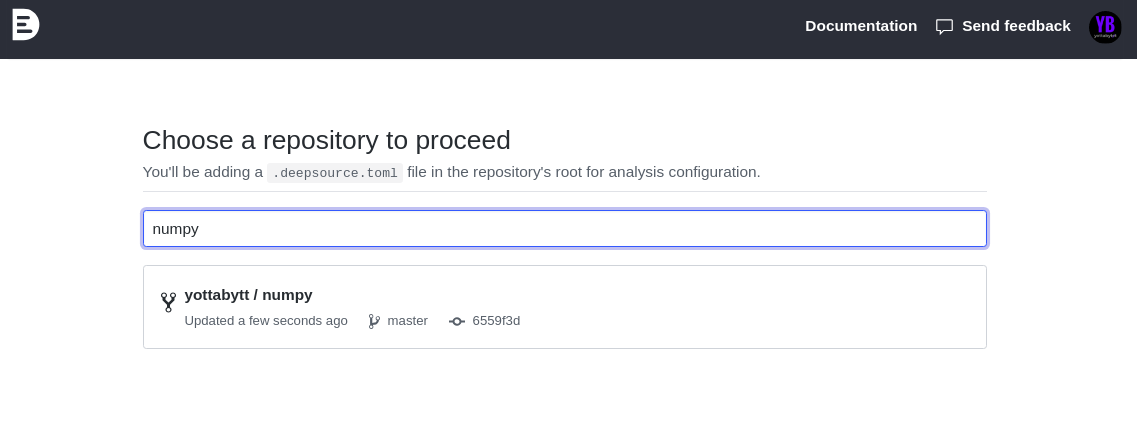}
\captionof{figure}{DeepSource - Choose repository. Search and select the repository. We chose the numpy repository (forked from the popular scientific computing package's repository).}
\label{fig:deepsource}
\end{figure}

\begin{figure}[H]
\centering
\includegraphics[width=\columnwidth,height=0.45\textheight,keepaspectratio]{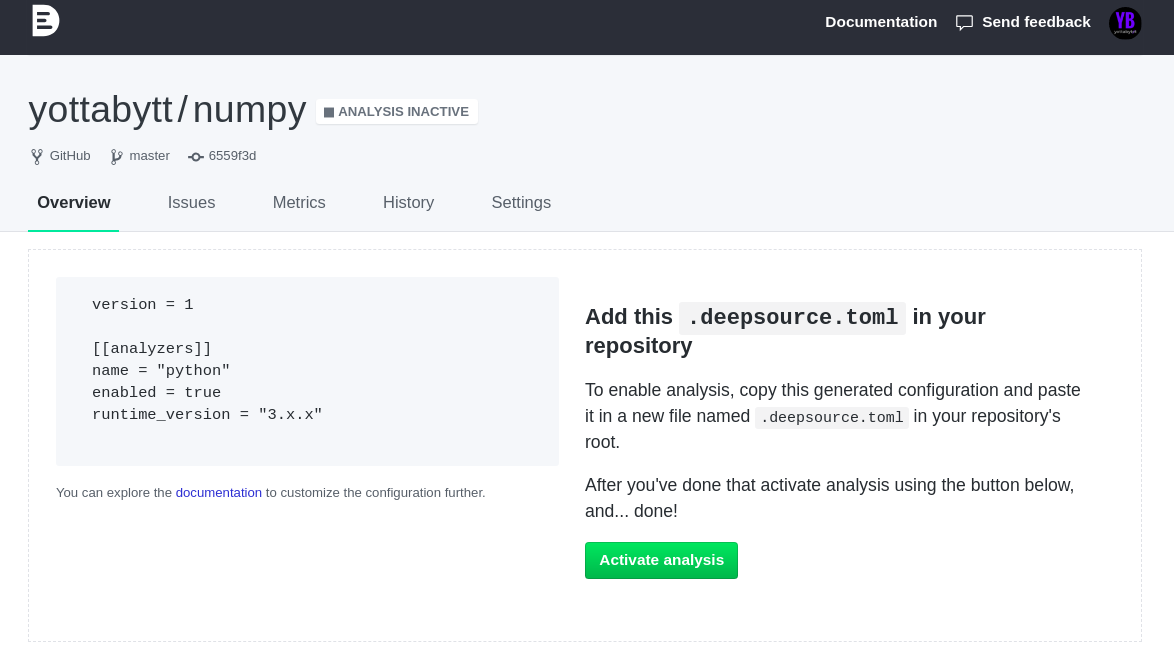}
\captionof{figure}{DeepSource - Activate analysis. Copy the .toml file as per the instructions above to make the analysis work.}
\label{fig:deepsource}
\end{figure}

\begin{figure}[H]
\centering
\includegraphics[width=\columnwidth,height=0.45\textheight,keepaspectratio]{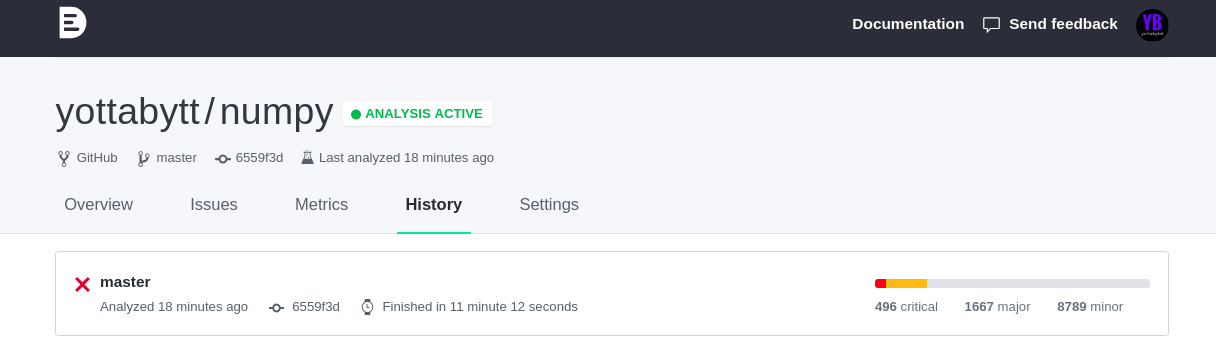}
\captionof{figure}{DeepSource - History tab that gives information about the current and previous analysis.}
\label{fig:deepsource}
\end{figure}

\section{Conclusion}

We have thus presented our empirical observations which we hope to be beneficial to both practitioners and researchers. We believe this work can serve as a bridge connecting the concepts that are popular in literature with the real world software tools which are both old and new. Above all, we suspect this work can give a quick and easy end-to-end understanding even for an absolute beginner in the field of \enquote{Technical Debt in Software Development}. However, we may have not addressed many of the tools which might be actually be more popular and useful. But still, we believe this can be a starter for works that includes them.

\section*{Acknowledgment}

We would like to thank \href{mailto:sean.barow@lattix.com}{Sean Barow} (Director of Sales, Lattix), for quickly accepting our request and granting us a limited time access to \enquote{Lattix}.

\bibliographystyle{unsrt}

\appendices
\section{} \label{sec:demo}
\subsection*{Architecture rule violations - A demonstration using Lattix} 

Lattix provides options to create multiple views that gives information about the project at different levels. Some of them are views for Dependency Structure Matrix(DSM) and Conceptual Architecture. DSM's can be very helpful in identifying the cross-cut communication between classes and methods belonging to different components (say packages). Also, one can view architecture rule violations right within the DSM. As none of the chosen projects had specified such rules, we were not able to witness it. However, here were try to witness it. 

\begin{figure}[H]
\centering
\includegraphics[width=\columnwidth,height=0.45\textheight,keepaspectratio]{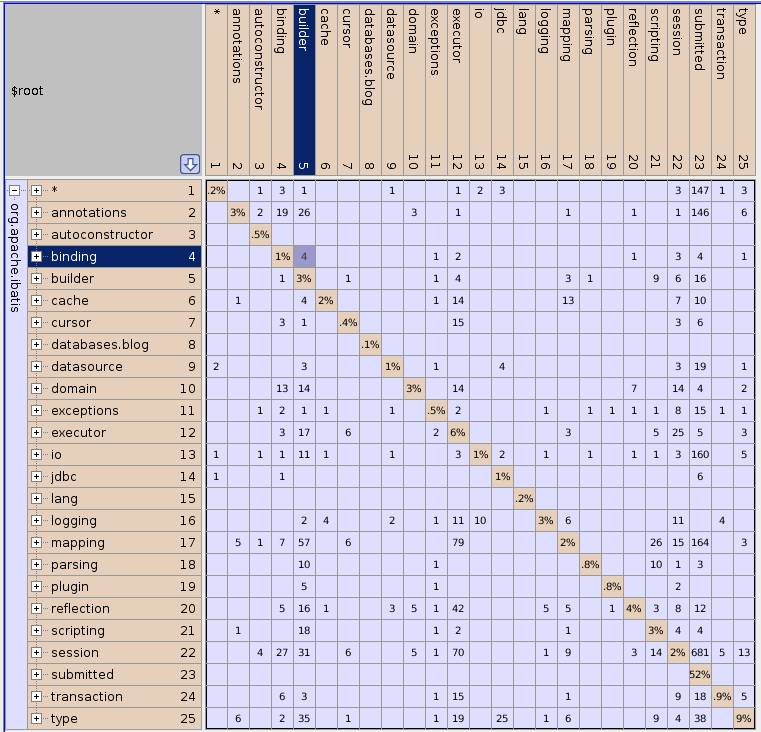}
\captionof{figure}{Dependency structure matrix (DSM) between top-level components from MyBatis}
\label{fig:dsm_before}
\end{figure}

\begin{figure}[H]
\centering
\includegraphics[width=\columnwidth,height=0.45\textheight,keepaspectratio]{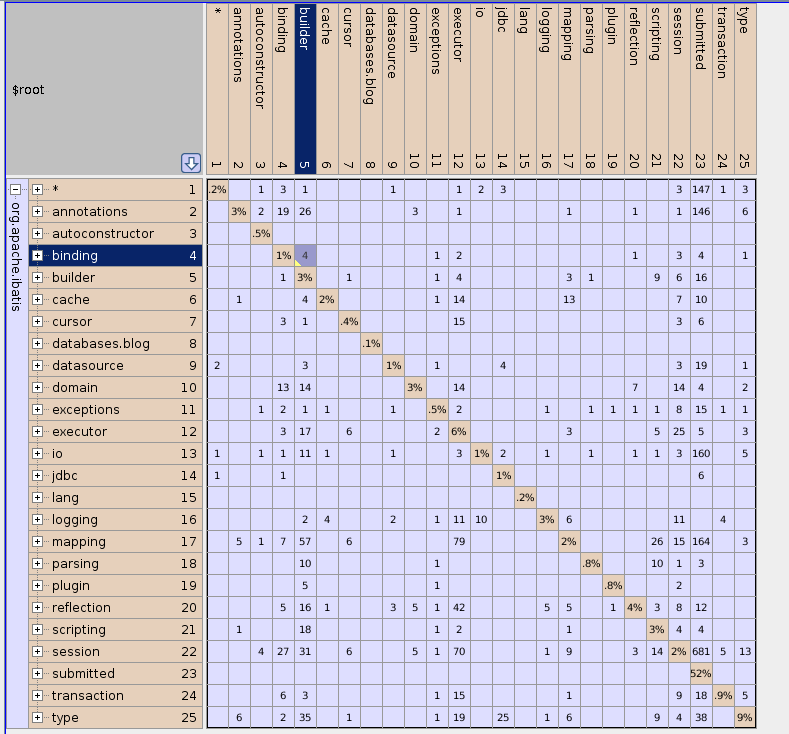}
\captionof{figure}{DSM when rules were enabled. Note the small yellow triangle at bottom left of the cell indicating a violation.}
\label{fig:dsm_after}
\end{figure}

\begin{figure}[H]
\centering
\includegraphics[width=\columnwidth,height=0.45\textheight,keepaspectratio]{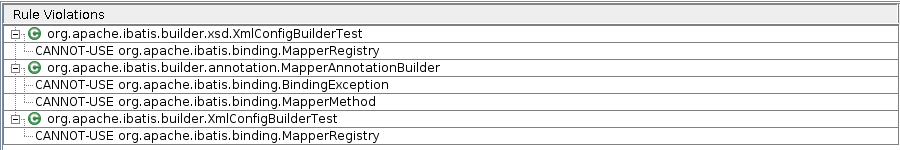}
\captionof{figure}{Rule violations view that gives further information about the violation that happens because of the dependencies.}
\label{fig:rule_violations}
\end{figure}

\begin{itemize}
\item Look at figure \ref{fig:dsm_before}, when there were no rules enabled and thus no violations for the project \enquote{MyBatis}
\item Lattix gives an option of enabling/disabling \enquote{Can-Use} and \enquote{Cannot-Use} rules between components. 
\begin{itemize}
\item Right Click on the cell which is highlighted in dark blue as shown in figure \ref{fig:dsm_before} which shows the dependency between the components \textit{org.apache.ibatis.binding} and \textit{org.apache.ibatis.builder}.
\item Select \enquote{Modify Rule} $\longrightarrow$ Select \enquote{Cannot-Use}
\item Now look at figure \ref{fig:dsm_after}, where you can see a small yellow triangle at bottom left of the cell indicating a violation.
\item More information about the violation is shown in a separate view as depicted in figure \ref{fig:rule_violations}.
\end{itemize}
\end{itemize}

\ifCLASSOPTIONcaptionsoff
  \newpage
\fi



%

%

\end{document}